\documentclass[global,twocolumn]{svjour}
\usepackage[utf8]{inputenc}
\usepackage{amsmath}
\usepackage{amsfonts}
\usepackage{amssymb}
\usepackage{graphicx}
\usepackage{dcolumn}
\usepackage{times}
\usepackage{tabularx}
\newcommand{\nth}[1]{\textsuperscript{#1}}
\begin{document}
\title{Laser-plasma accelerator based single-cycle attosecond undulator source}
\author{Z. Tibai\inst{1}\and Gy.\ T\'oth\inst{2}\and A. Nagyv\'aradi\inst{3}\and A. Sharma\inst{4}\and M. I. Mechler\inst{2}\and J. A. F\"ul\"op\inst{2,5}\and G. Alm\'asi\inst{1,2}\and J. Hebling\inst{1,2,5}}
\institute{Institute of Physics, University of Pécs, 7624 P\'ecs, Hungary\and
MTA-PTE High Field Terahertz Research Group, 7624 P\'ecs, Hungary\and
Faculty of Engineering and Information Technology, University of P\'ecs, 7624 P\'ecs, Hungary\and
ELI-ALPS, Szeged 6720, Hungary\and 
Szent\'agothai Research Centre, University of P\'ecs, 7624 P\'ecs, Hungary\\\email{tibai@fizika.ttk.pte.hu}}
\maketitle
\begin{abstract}
Laser-plasma accelerators (LPAs), producing high-quality electron beams, provide an opportunity to reduce the size of free-electron lasers (FELs) to only a few meters. A complete system is proposed here, which is based on FEL technology and consists of an LPA, two undulators, and other magnetic devices. The system is capable to generate carrier-envelope phase stable attosecond pulses with engineered waveform. Pulses with up to~60~nJ energy and 90 to~400~attosecond duration in the 30 to 120~nm wavelength range are predicted by numerical simulation. These pulses can be used to investigate ultrafast field-driven electron dynamics in matter. 
%%Missing OCIS; ez kell vagy keywords?
\end{abstract}
\section{Introduction}\label{sec:intro}
The time-resolved study of electron dynamics in atoms, mol\-e\-cules, and solids require attosecond-scale temporal resolution \cite{HentschelNat2001,KienbergerNat2004,SchaferPRL2004,JohnssonPRL2005,KrauszRMP2009}. A suitable tool is provided by isolated attosecond pulses in the extreme ultraviolet (EUV) spectral range, produced by high-harmonic generation (HHG) of waveform-controlled few-cycle laser pulses. The recent demonstration of single-cycle isolated attosecond pulses \cite{KrauszRMP2009} may open the way to a new regime in ultrafast physics, where the strong-field electron dynamics is driven by the electric field of the attosecond pulses rather than by their intensity profile. Intense EUV pulses can also be generated in free-electron lasers (FELs), which are expensive large-scale facilities relying on linear accelerators (LINACs). In recent years, many ideas have been proposed to achieve ultrashort pulses in FELs \cite{EmmaPRL2004,ZholentsPRST2005,DunningPRL2013,GarciaPRAB2016,TanakaPRL2015,SansoneSci2006}. However, the limited temporal coherence of the radiation remained a serious drawback for many applications.

Laser-plasma based electron accelerators (LPAs) \cite{TajimaPRL1979,ManglesNature2004,GeddesNature2004,FaureNature2004,MalkaNP2008,EsareyRMP2009,WangNC2013} can be a cost-effective alternative for large-scale LINACs, allowing a tremendous reduction of the accelerator size. It is possible to reach electron energies up to the GeV level over acceleration distances of only a few cm in a wakefield LPA with parameters comparable to (and in some aspects even better than) conventional LINAC sources \cite{EsareyRMP2009,WangNC2013,LeemansNP2006}. The electron beam transversal size in a LPA can be more than two orders of magnitude smaller ($\sim1\mu$m) than typical for LINACs. The peak current in a LPA can be one order of magnitude higher than in a conventional LINAC. Such unique properties of LPA electron pulses offer the potential to construct FELs with a significantly reduced \emph{overall} dimension and cost, thereby enabling a widespread proliferation of FEL sources to small-scale university labs. Owing to these advantages, numerous schemes have been proposed to use LPA electrons to drive FEL sources \cite{CoupriePPCF2016,TilborgAIPCP2017,LiuPRAB2017,NakajimaNP2008,FuchsNP2009,CouprieJPB2014,CouprieJMO2016,FengOE2015}.

Remarkably, none of the previously mentioned techniques, whether HHG-, LPA-, or LINAC-based, could so far produce waveform-engineered CEP-controlled attosecond pulses. Therefore, previously we proposed a robust method for producing such EUV pulses using conventional LINAC-based FEL technology \cite{TibaiPRL2014,TothOL2015,TibaiArxiv2016}. Motivated by the exciting recent progresses in high-power lasers and LPAs delivering high-quality electron beams, here we investigate the feasibility and the prospects of adopting our previous concept to be driven by an LPA, rather than a LINAC. A detailed conceptual study is presented, based on numerical simulations, for an entirely laser-driven waveform-engineered CEP-controlled attosecond pulse source.
\begin{figure*}[!htb]
\includegraphics[width=\textwidth]{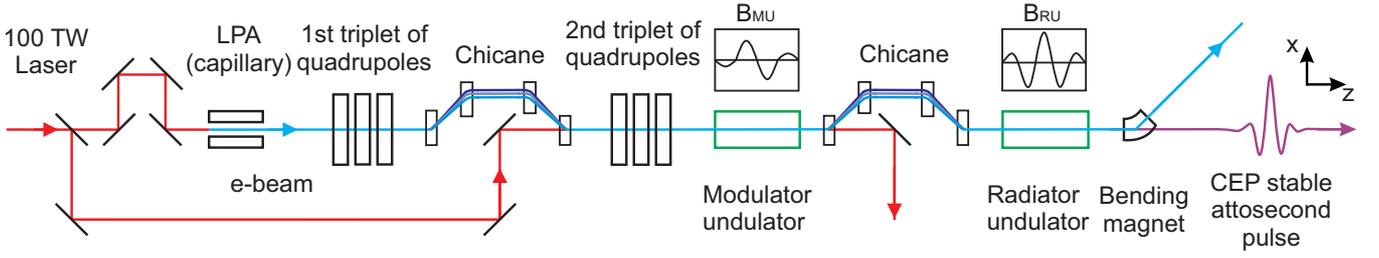}
\caption{\label{fig:1}Scheme of the proposed LPA-based setup.}
\end{figure*}

\section{The proposed setup}\label{sec:setup}
The proposed setup for an LPA-based waveform-engineered attosecond pulse source is shown in Fig.~\ref{fig:1}. Laser pulses of at least 100 TW-power are split into two separate beams. The main part is focused into a gas-filled capillary for producing a relativistic electron beam \cite{EsareyRMP2009,LeemansNP2006}. After the LPA, the electron beam is sent through the first triplet of quadrupoles to reduce its divergence. Subsequently, a chicane reduces the slice energy spread of the beam. The following second triplet of quadrupoles is used to focus the electron beam into the radiator undulator (RU), which is placed further downstream in the setup. The second quadrupole triplet is followed by the modulator undulator (MU), where the electron bunch is overlapped with the smaller, 20-TW portion of the laser beam. Here, the interaction between the electrons, the magnetic field of the MU, and the electromagnetic field of the modulator laser introduces a spatially periodic energy modulation of the electrons. The electrons propagate through a second chicane and their energy modulation leads to the formation of a train of nanobunches (ultrathin electron layers), separated by the modulator laser wavelength. (The second chicane is used to reduce the propagation distance where the shortest nanobunches are formed.) The nanobunched electron beam then passes through the RU, consisting of a single or a few periods, and creates CEP-stable attosecond pulses.

\section{Transport and manipulation of the electron beam}\label{sec:trans}
The General Particle Tracer (GPT) numerical code \cite{GPT} was used for the simulation of the electron beam transport from the capillary to the RU. Due to limitations of computational capacity, macroparticles consisting of about 4100 electrons were considered, rather than individual electrons. The initial electron bunch parameters were chosen according to feasible parameters for LPA-generated electron beams \cite{EsareyRMP2009,LeemansNP2006} and are listed in Table~\ref{tab:1}. We performed a series of simulation runs, during which the electron bunch length was kept constant for each run, but the initial spatial distribution of the electrons was random. We note that particle-in-cell (PIC) simulations \cite{BussmannSC2013} can be used to implement more detailed predictions on the initial electron beam parameters, which then can serve as input for the particle-tracking calculations. However this is out of scope for the present work aiming at assessing the feasibility of the LPA-based CEP-stable attosecond source, rather than giving a complete in-depth numerical design and optimization study.
\begin{table}[!htb]
\caption{\label{tab:1}Parameters of the electron beam originating from an LPA, used as initial values in the simulations.}
\begin{tabularx}{\columnwidth}{XD{,}{~}{8}}\hline
Parameter&\text{Value}\\\hline
Energy ($\gamma_0$)&2000,\\
Energy spread ($\sigma_{\gamma_0}$)&1.0,\%\\
Normalized emittance ($\gamma_0\varepsilon_{x,y}$)&1.0,\textrm{mm~mrad}\\
Transversal size ($\sigma_{x0}=\sigma_{y0}$)&1.0,\mu\textrm{m}\\
Length ($\sigma_{z0}$)&1.0,\mu\textrm{m}\\
Charge&50,\textrm{pC}\\\hline
\end{tabularx}
\end{table}
Efficient generation of radiation is possible only if the (micro- or nano) bunch length is shorter than half of the radiation wavelength. If the Coulomb interaction between the electrons in the bunch can be neglected, the following analytical formula can be derived for the FWHM of the electron density of the micro- or nanobunch:
\begin{equation}
\Delta b_0=\frac{\lambda_l\sigma_{\gamma}}{2\Delta\gamma_{\text{MU}}},\label{eq:b0}
\end{equation}
where $\lambda_l$ is the wavelength of the modulator laser, $\sigma_{\gamma}$ is the energy spread of the nanobunch and $\Delta\gamma_{\text{MU}}$ is the induced energy gain in the MU \cite{HemsingRMP2014,Zholents2PRST2005}. According to this equation, shorter nanobunches can be achieved by using an electron beam with smaller energy spread and stronger energy modulation.

The energy spread of an LPA generated electron beam is typically much larger (1-10\%) than that of a LINAC (typically $<$0.05\%). Therefore, a reduction of the slice energy spread is necessary, which can be accomplished by the stretch of the beam or utilizing a transverse-gradient undulator (TGU) as a radiator \cite{HuangPRL2012}. In our setup we used the first chicane to decompress the beam longitudinally. The chicane consisted of four identical dipole magnets with parameters as listed in Table~\ref{tab:2}. The slice energy spread of the electron beam is reduced at the end of the chicane according to the formula \cite{CouprieJPCS2013}
\begin{equation}
\sigma_{\gamma z}=\sigma_{\gamma 0}\frac{\sigma_{z0}}{\sigma_{zz}}=\sigma_{\gamma 0}\frac{\sigma_{z0}}{\sqrt{\sigma_{z0}^2+\left(R_{56}\frac{\sigma_{\gamma 0}}{\gamma_0}\right)^2}},\label{eq:sgz}
\end{equation}%Eq1
where $\sigma_{zz}$ is the length of the electron beam after the chicane, $R_{56}=2\theta_0^2\left(D+2/3L\right)$, $D$ is the distance between 1\nth{st} (3\nth{rd}) and 2\nth{nd} (4\nth{th}) dipoles, $L$ is the distance between 2\nth{nd} and 3\nth{rd} dipoles, and $\theta_0$ is the deviation angle at the dipoles. Hence, the electron bunch can be easily lengthened by adjusting the chicane strength ($R_{56}$). The slice energy spread of the beam is reduced from 1\% to 0.1\% with $R_{56}=1$~mm, and the bunch is lengthened to about 10~$\mu$m (33~fs).

\begin{table}[tb]
\caption{\label{tab:2}Parameters of the first chicane.}
\begin{tabularx}{\columnwidth}{XD{,}{~}{2}}\hline
Parameter&\text{Value}\\\hline
Dipole length&15,\text{cm}\\
Gap&1,\text{cm}\\
Dipole strength&0.8,\text{T}\\
Distance between 1\nth{st} (3\nth{rd}) and 2\nth{nd} (4\nth{th}) dipoles ($D$)&30,\text{cm}\\
Distance between 2\nth{nd} and 3\nth{rd} dipoles ($L$)&20,\text{cm}\\\hline
\end{tabularx}
\end{table}
The modulator laser pulses are used for driving the nanobunch formation. As mentioned in the previous section, these laser pulses are provided by splitting off a small portion of the main beam. Our setup is very advantageous, because one common laser source creates the electron beam and the modulator laser beam, too. Therefore, the synchronization between the electron beam and the modulator laser pulses can be easily solved. Today, numerous 100-TW class laser systems are operating all around the world. For the split-off modulator laser pulses 20 TW peak power was assumed (Table \ref{tab:3}). The MU period $\lambda_{MU}$ was chosen to satisfy the well-known resonance condition 
\cite{Wiedemann2003} at $K_{\text{MU}}=2.0$,
 where $K_{\text{MU}}=(eB_{\text{MU}}\lambda_{\text{MU}})/\allowbreak (2\pi m_e c)$. Here, $e$ is the electron charge, $B_{\text{MU}}$ is the magnetic field amplitude, $m_e$ is the electron rest mass, and $c$ is the speed of light.
A double-period MU was assumed with antisymmetric design of $-1/4$, $3/4$, $-3/4$, and $1/4$ relative magnetic field amplitudes. According to our calculations, the maximum energy modulation in the MU is $\Delta\gamma_{\text{MU}}\approx140$. We note that the coherent synchrotron radiation effect may reduce the quality of the electron beam inside the MU and the chicanes. Although our estimations predicted that this effect is negligible, a more detailed investigation is planned using the SIMPLEX \cite{TanakaJSR2015} and the GENESIS \cite{ReicheNIMP1999} codes.

\begin{figure}[tb]
\includegraphics[width=\columnwidth]{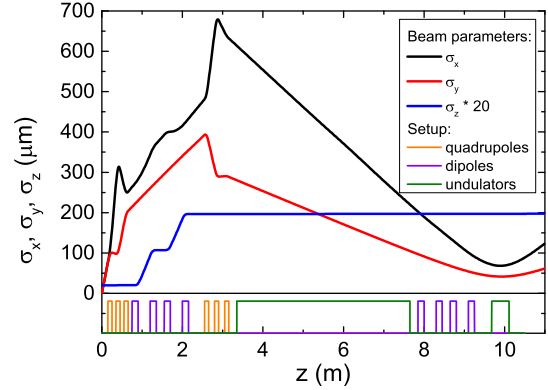}
\caption{\label{fig:2}Variation of the transversal and longitudinal electron bunch sizes along the propagation through the magnetic components of the setup. The size and position of each component is indicated in the bottom part of the figure.}
\end{figure}

\begin{table}[tb]
\caption{\label{tab:3}Parameters of the modulator laser and the MU.}
\begin{tabularx}{\columnwidth}{XD{,}{~}{2}}\hline
Parameter&\text{Value}\\\hline
Laser wavelength ($\lambda_l$)&800,\text{nm}\\
Laser peak power&20,\text{TW}\\
Laser beam waist&1,\text{mm}\\
Laser beam Rayleigh length&3.9,\text{m}\\
MU undulator parameter ($K_{\text{MU}}$)&2.0\\
MU period length ($\lambda_{\text{MU}}$)&2.15,\text{m}\\
MU magnetic field amplitude ($B_{\text{MU}}$)&0.01,\text{T}\\\hline
\end{tabularx}
\end{table}
The location of the temporal focal point (where the nano\-bunch length becomes the shortest) was controlled with the second chicane. The total charge in a single nanobunch is about $\sim0.6$~pC and its length is as short as 27~nm (FWHM). The RU was placed around the temporal focal point. The transversal focusing of the electron beam at the position of the RU was achieved with the two permanent-magnet quadrupole triplets with parameters as shown in Table~\ref{tab:4}. The gradients of them were optimized with GPT and a self-developed code. The calculated variations of transversal and longitudinal sizes of the whole electron bunch along the propagation direction are shown in Fig.~\ref{fig:2}. According to our calculations the transversal size of the electron-bunch at the RU in the $x$ and $y$ directions are 42~$\mu$m and 68~$\mu$m, respectively. As shown in Fig.~\ref{fig:2}, the variation of the transversal size of the bunch inside the RU is negligible.

\begin{table}[tb]
\caption{\label{tab:4}Parameters of the quadrupoles.Three different values were used for $K_{\text{RU}}$, and correspondingly, for $\lambda_{\text{RU}}$.}
\begin{tabularx}{\columnwidth}{Xr@{\;}l}\hline
Parameters&\multicolumn{2}{c}{\text{Value}}\\\hline
Distance between the capillary and the 1\nth{st} quadrupole&15&\text{cm}\\
1\nth{st} quadrupole triplet gradients&197 / 177 / 73&\text{T/m}\\
1\nth{st} chicane dipole field&0.8&\text{T}\\
2\nth{nd} quadrupole triplet gradients&47 / 51 / 6&\text{T/m}\\
2\nth{nd} chicane dipole field&0.011&\text{T}\\
RU undulator parameter ($K_{\text{RU}}$)&0.5, 0.8, \text{and} 1.2&\\
RU undulator period length ($\lambda_{\text{RU}}$)&43, 36, \text{and} 28&\text{cm}\\\hline
\end{tabularx}
\end{table}

\section{Single cycle attosecond pulse generation}\label{sec:ssapg}
The temporal shape of the attosecond EUV pulses emitted by the extremely short nanobunches in the RU were calculated at a plane positioned 8~m behind the RU center. EUV waveform engineering is conveniently enabled by designing the magnetic field distribution of the RU. In the present case the magnetic field of the RU was defined as
\begin{equation}
B_{\text{RU}}=\left\{
\begin{array}{cc}
\displaystyle B_{\text{RU0}}{\rm e}^{-\frac{z^2}{2\sigma^2}}\cos\left(\frac{2\pi}{\lambda_{\text{RU}}}z+\varphi\right),&\text{if }-\frac{W}{2}<z<\frac{W}{2}\\
0&\text{otherwise}
\end{array}\right.,\label{eq:BRU}
\end{equation}%Eq2
where $B_{\text{RU0}}$ is the peak magnetic field, $z$ is the length along the direction of the nanobunch propagation, $\sigma$ is the standard deviation of the Gaussian envelope, $\varphi$ is the CEP and $W$ is the length of the RU. In each simulation series, the undulator period was adjusted to give the desired central wavelength for the radiation (20~nm, 30~nm, 40~nm, 60~nm, 80~nm, 100~nm, and 120~nm), according to the resonance equation. The further undulator parameters were set to $\sigma=1.5\times\lambda_{\text{RU}}$, $W=2.5\times\lambda_{\text{RU}}$, and $\varphi=0$. (Similar magnetic field distribution has been demonstrated in the experiment of Kimura et al. \cite{KimuraPRST2004}.)
\begin{figure*}[tb]
\includegraphics[width=\columnwidth]{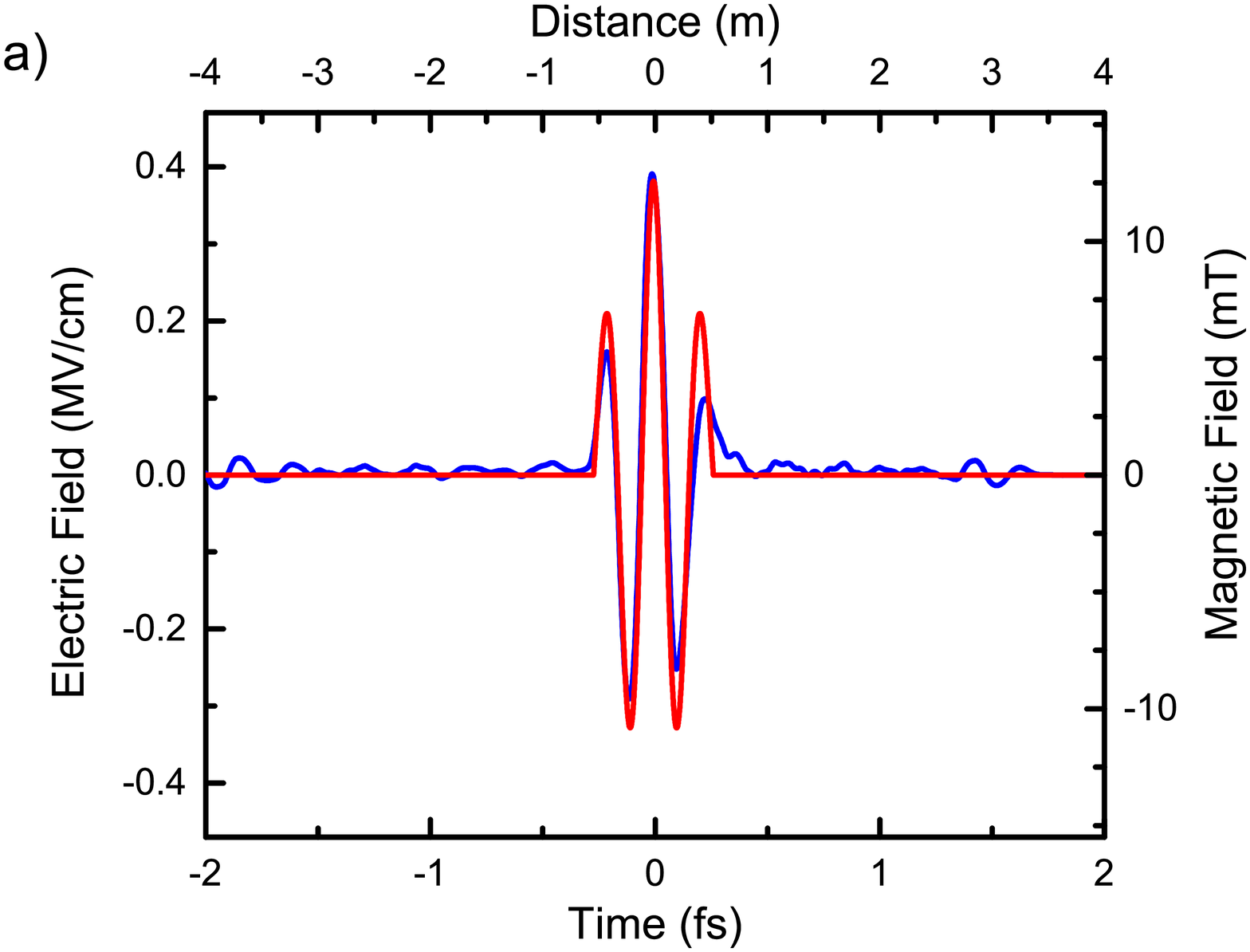}
\includegraphics[width=\columnwidth]{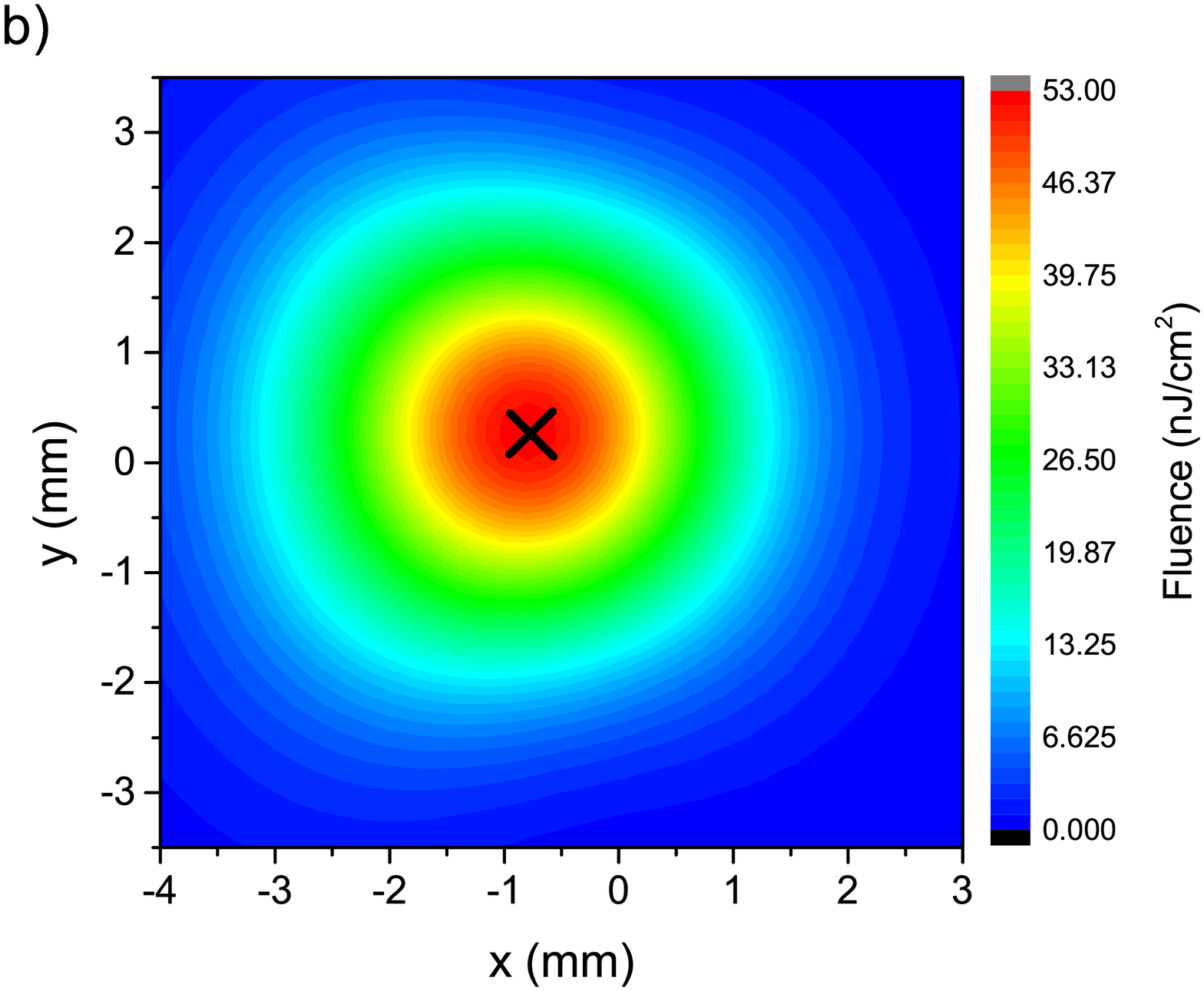}
\caption{\label{fig:3}(a) Example of a CEP-controlled EUV waveform for 60~nm radiation wavelength and $K_{\text{RU}}=0.5$. (b) The corresponding spatial beam profile.}
\end{figure*}
\begin{figure*}[tb]
\includegraphics[width=\columnwidth]{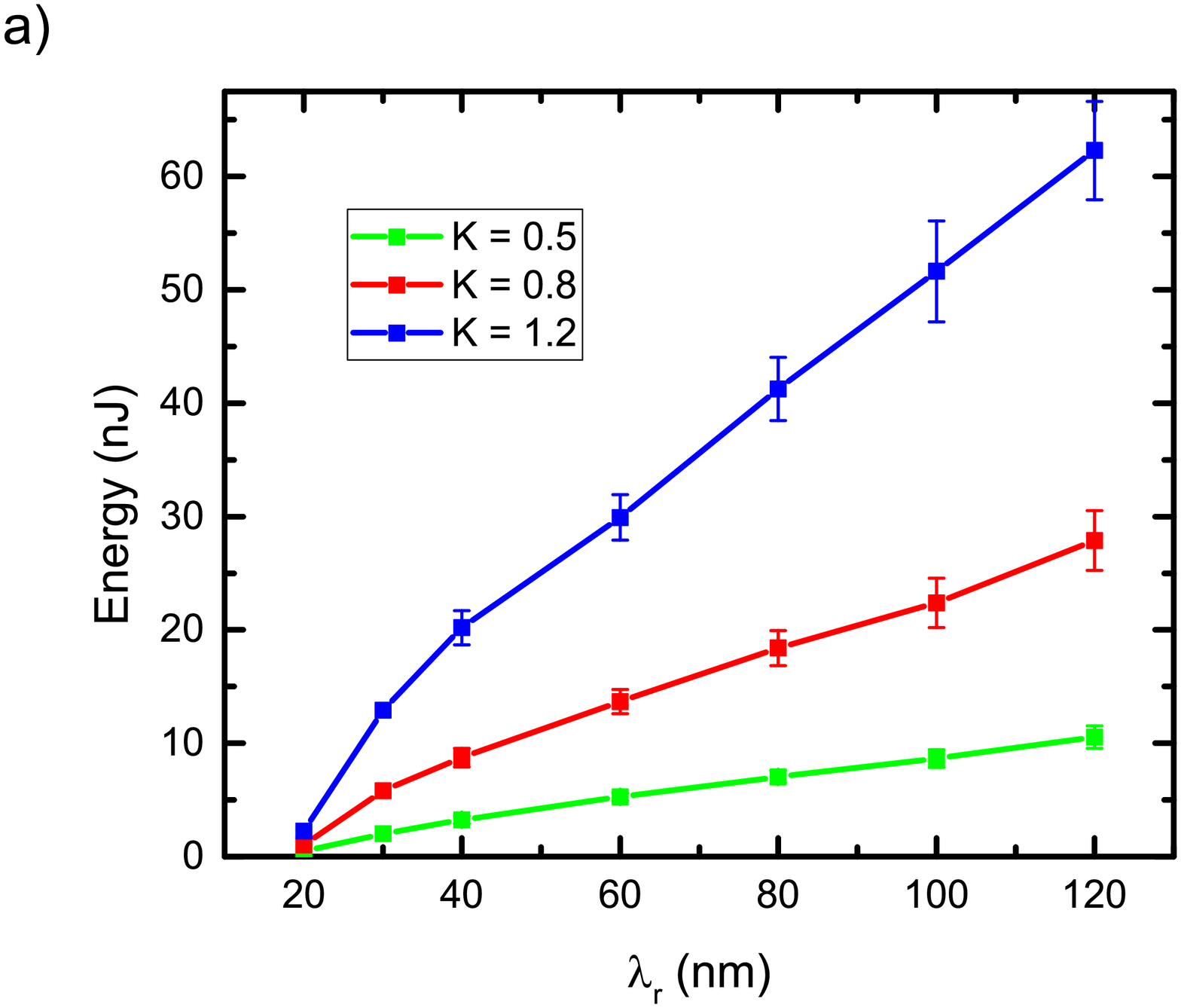}
\includegraphics[width=\columnwidth]{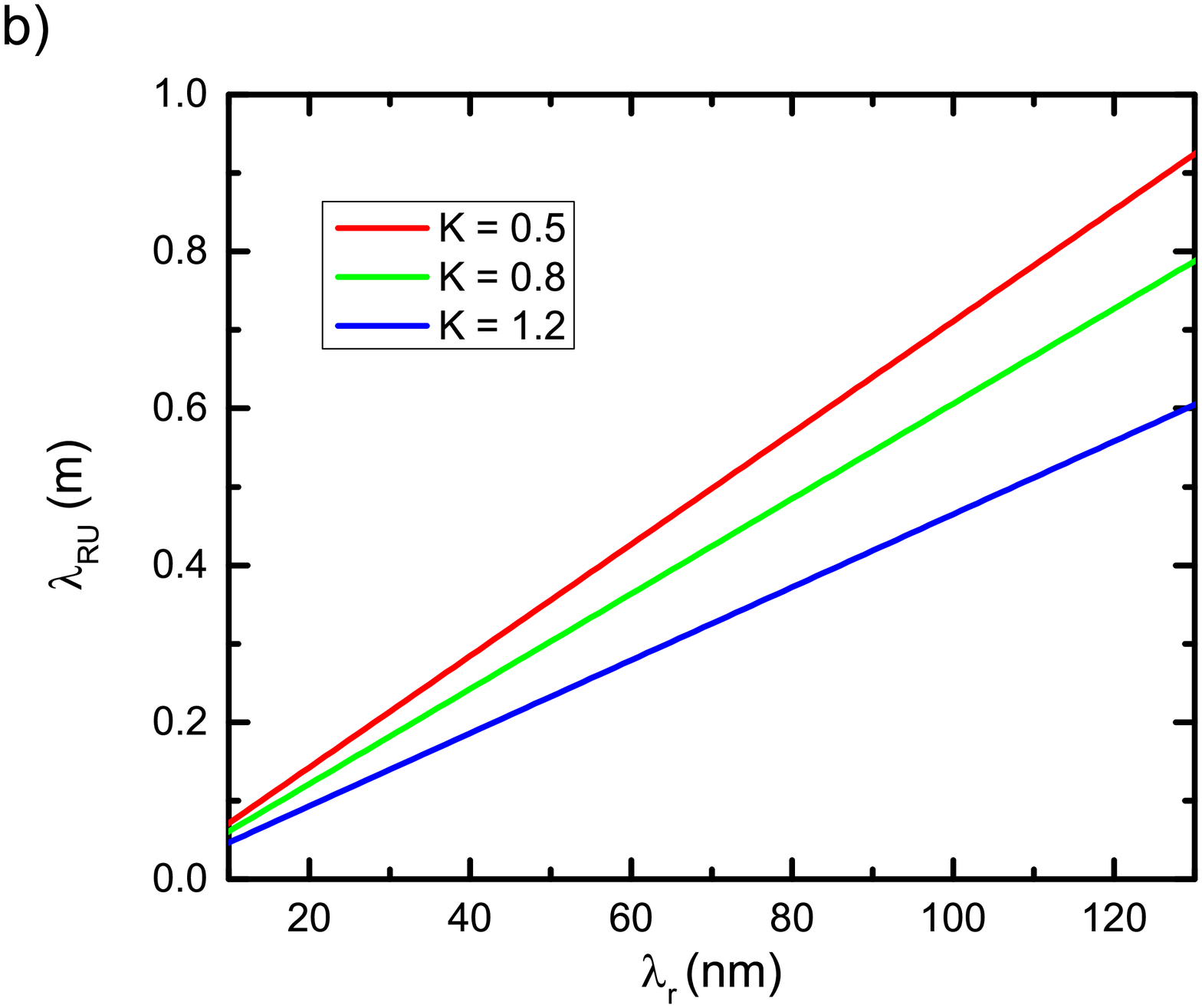}
\caption{\label{fig:4}The EUV pulse energy (a) and the period of the RU (b) as functions of the radiation wavelength $\lambda_{\text{r}}$ for different RU undulator parameters.}
\end{figure*}

We used the following handbook formula to calculate the electric field of the radiation generated in the RU \cite{Jackson2007}:
\begin{equation}
\vec{E}(t,\vec{r})=\sum\left[\frac{q\mu_0}{4\pi}\frac{\vec{R}\times\left((\vec{R}-R\vec{\beta})\times\dot{\vec{v}}\right)}{(R-\vec{R}\cdot\vec{\beta})^3}\right]_{\text{ret}},\label{eq:E}
\end{equation}%Eq3
where $\mu_0$ is the vacuum permeability, $q$ is the macroparticle charge, $\vec{R}$ is the vector pointing from the position of the macroparticle at the retarded moment to the observation point, $\vec{v}$ is the velocity of the macroparticle, $\vec{\beta}=\vec{v}/c$, and $c$ is the speed of light. The summation is for all macroparticles. During the radiation process the position, velocity, and acceleration of the macroparticles were traced numerically by taking into account the Lorentz force of the magnetic field of RU. The Coulomb interaction between the macroparticles was neglected during the undulator radiation process, because the transversal electron motion is by four orders of magnitude larger than the motion generated by Coulomb interaction \cite{TibaiPRL2014}.

Figure~\ref{fig:3}a displays one example of the simulated waveform of the generated attosecond pulse (blue curve) and Figure~\ref{fig:3}b the corresponding beam profile for 60~nm radiation wavelength and $K_{\text{RU}}=0.5$ (the cross symbol marks the location where the waveform in Fig.~\ref{fig:3}a was sampled). As it is shown in Fig.~\ref{fig:3}a, the waveform of the generated attosecond pulse resembles the magnetic field of the RU (red curve) \cite{TibaiPRL2014}. At other wavelengths the shapes of the attosecond pulses are nearly identical to the shape shown in Fig.~\ref{fig:3}a (blue curve). Besides the CEP stability of the EUV pulses, another important advantage of this setup is that the EUV pulse CEP can be controlled (set) by the magnetic field distribution of the RU \cite{TothNIMP2016}. Consequently, attosecond pulses with both single- and multi-cycle waveform can be generated with this technique.

The EUV pulse energy as function of the radiation wavelength ($\lambda_{\text{r}}$) in the range of 20~nm to 120~nm is shown in Fig.~\ref{fig:4}a, with three different RU undulator parameter values of 0.5, 0.8, and 1.2. These values correspond to 13 mT, 24 mT and 46 mT peak magnetic fields for $\lambda_{\text{r}} = 60$ nm. Larger EUV pulse energy is obtained with larger $K_{\text{RU}}$, because the energy of the pulse is proportional to the square of $K_{\text{RU}}$ value. Fig.~\ref{fig:4}a contains the results of about 10 numerical simulation runs for every parameter set, determining also the error bars. The undulator parameter can be set to the desired value by adjusting the magnetic field amplitude. The radiation wavelength, given by the resonance condition, can be set by the choice of the RU period $\lambda_{\text{RU}}$. The period of the RU as function of the radiation wavelength $\lambda_{\text{r}}$ for different RU undulator parameters is shown in Fig.~\ref{fig:4}b. Our simulations predict attosecond pulses with up to 10~nJ, 25~nJ, and 60~nJ energy for undulator parameters of 0.5, 0.8, and 1.2, respectively (Fig.~\ref{fig:4}a). These energies are sufficient for many applications. Importantly, the LPA-based scheme does not require a large-scale accelerator facility and can be affordable for smaller laboratories.

\section{Conclusion}
A robust method for the efficient generation of CEP-stable single-cycle attosecond pulses was proposed, which utilizes a laser-plasma based electron accelerator, a modulator undulator, and a radiator undulator. The laser pulses used to drive the electron source and the nanobunching can be derived from the same laser, easily enabling the precise synchronization between the electrons and the modulating laser field. The waveform of the attosecond pulses can be engineered by the choice of the magnetic field distribution in the radiator undulator. A conceptual design study was presented, including also a combination of magnetic devices for electron beam transport and manipulation. The generation of single-cycle attosecond pulses in the EUV spectral range with up to 60~nJ energy was predicted by numerical simulations. The results clearly show that the previously proposed LINAC-based scheme can be adopted to an entirely laser-driven one, thereby enabling to shrink the size of the system from hundreds-of-meters to only a few meters, and a cost-effective implementation in small-scale laboratories. The source is suitable to deliver pump pulses in pump-probe measurements and in time- and CEP-resolved measurements with $\sim$100-as resolution.

\begin{acknowledgement}
This work is partially supported by European Cluster of Advanced Laser Light Sources (EUCALL) project which has received funding from the European Union's Horizon 2020 research and innovation programme under grant agreement No. 654220. Financial support from Hungarian Scientific Research Fund (OTKA) grant No. 113083 is acknowledged. JAF acknowledges support from J\'anos Bolyai Research Scholarship (Hungarian Academy of Sciences). The present scientific contribution is dedicated to the 650\nth{th} anniversary of the foundation of University of P\'ecs, Hungary.
\end{acknowledgement}
\bibliographystyle{unsrt}
\bibliography{TZbib}
\end{document}